\documentstyle[12pt]{aipproc}

\newcommand{\eq}[1]{Eq.~(\ref{#1})}

\def\beq{\begin{equation}}
\def\eeq{\end{equation}}
\def\beqa{\begin{eqnarray}}
\def\eeqa{\end{eqnarray}}

\newcommand{\EQ}{\begin{equation}}
\newcommand{\EN}{\end{equation}}
\newcommand{\bea}{\begin{eqnarray}}
\newcommand{\ena}{\end{eqnarray}}

\renewcommand{\a}{\alpha}

\newcommand{\ve}{\varepsilon}

\begin{document}

\title{String derivation of two-loop Feynman diagrams\footnote{Talk
presented at ``Beyond the Standard Model $V$'', Balholm, Norway, April
1997. The work described here is part of an ongoing collaboration 
with P. Di Vecchia, A. Lerda and R. Marotta. Preprint 
{\bf NORDITA--97/57 P}.}}

\author{Lorenzo Magnea$^{\diamond}$ and Rodolfo Russo$^{\ddagger}$}

\address{$^{\diamond}$NORDITA\thanks{On leave from Universit\`a di
Torino, Italy.}\\
Blegdamsvej 17, DK--2100 Copenhagen \O, Denmark\\
$^{\ddagger}$Dipartimento di Fisica, Politecnico di Torino\\
Corso Duca degli Abruzzi 24, I--10129 Torino, Italy\\}

\maketitle

\begin{abstract}
We briefly review the technology involved in extracting the
field--theory limit of multiloop bosonic string amplitudes, and we
apply it to the evaluation of simple two-loop diagrams involving 
scalars and gauge bosons.

\end{abstract}

\section*{Introduction}

The discovery of new physics beyond the Standard Model (SM) is most 
likely to come from experiments performed at the highest available 
energy; in fact, it is conceivable that hints of such a discovery 
have already been seen~\cite{lm_hera}. It is, however, very difficult 
to substantiate a discovery claim, if the SM predictions on the matter 
are not sufficiently sharp. A competing explanation of the new
phenomena may be constructed within the boundaries of the Standard 
Model, by tuning or stretching some of its parameters~\cite{lm_pdf}. 
It is clear that sharpening our techniques to make more precise
SM--based predictions on the outcome of high energy experiments
is a necessary tool for the discovery of new physics. 

Our goal here is to give a progress report on the development of some
of these techniques, and we will focus on the calculation of multiloop
scattering amplitudes in field theory using string theory as a
calculational tool. The method has been in use for several 
years~\cite{lm_all}, and recently considerable progress has 
been made towards its extension to more than one loop~\cite{lm_us}.
Here we would like to give a flavor of how these calculations are
performed, by reviewing some of the general features of the method, 
and then discussing two simple examples, one in Yang--Mills theory, 
the second in a scalar field theory.

By way of introduction, let us briefly recall some of the general
features of string--inspired techniques. The field theory limit of 
a string amplitude is obtained by taking the string tension 
$T = 1/(2\pi\alpha')$ to infinity, decoupling all massive string
modes; in this limit, the string world--sheet degenerates into a
graph, and the only corners of string moduli space contributing to 
the field theory result are those where the integrand of the string 
amplitude exhibits a singular behavior; string moduli are then related
to Schwinger proper times in field theory, and string-derived
amplitudes are expressed as those derived using the world-line
formalism in field theory~\cite{lm_sss}. 
The main advantages of resorting to a theory as complex as string
theory to perform Feynman diagram calculations are discussed in detail
in~\cite{lm_all}: concisely, the calculation of a Yang--Mills
scattering amplitude is reduced to roughly the size of the calculation
of the same amplitude in a scalar theory; loop momentum integrals are 
already performed, so that helicity methods can be readily employed,
and the result for a set of Feynman diagrams of a given topology 
is presented directly as a Schwinger--parameter integral, bypassing
the tensor algebra associated with the propagation of spin--$1$ 
particles.

Because of the limited space available, we cannot give details of the
derivation of the amplitudes from the string `master formula'. We will
just briefly review the general formalism, and hopefully clarify some
aspects of the method by discussing two simple examples.

\section*{General features of the method}

Let us begin by recalling the general expression for the color--ordered
$h$--loop $M$--gluon planar amplitude in the open bosonic
string~\cite{lm_us},
\beqa
A^{(h)}_P & = & C_h {\cal N}_0^M \int [dm]_h
\frac{\prod_{i=1}^M dz_i}{dV_{abc}} \prod_{i<j}
\left[\frac{\exp\left({\cal G}^{(h)}(z_i,z_j)\right)}{
\sqrt{V'_i(0)\,V'_j(0)}}\right]^{2\a' p_i\cdot p_j}
\label{lm_hmast} \\
& \times & \exp \left[\sum_{i \not= j} \left(
\sqrt{2\a'} p_j\cdot\ve_i
\,\partial_{z_i} {\cal G}^{(h)}(z_i,z_j)
+ \, {1\over 2}\ve_i\cdot\ve_j
\,\partial_{z_i}\partial_{z_j}
{\cal G}^{(h)}(z_i,z_j)\right)\right]~,
\nonumber
\eeqa
where only terms linear in each polarization should be kept, and we
omitted the color factor $N^h\,{\rm Tr}(\lambda^{a_1}\cdots 
\lambda^{a_M})$. The amplitude for the scattering of $M$ scalars at
$h$ loops is obtained by omitting the second line and inserting a
factor $\prod_i [V_i'(0)]^{(-1)}$ in the measure of integration.
The fundamental ingredients of \eq{lm_hmast} are the bosonic Green 
function ${\cal G}^{(h)}(z_i,z_j)$ (the correlator of two scalar
fields on the $h$-loop string world sheet), and the measure of 
integration on moduli space $[dm]_h$. 
Both these quantities depend only on the genus $h$ of the
surface. Expressions for the Green function and the measure at 
two loops, in the field theory limit, will be given below. All 
geometric quantities are expressed in the Schottky parametrization 
of Riemann surfaces, which is obtained by cutting and gluing circles
on the Riemann sphere. Each pair of circles (each loop) is defined by
a projective transformation, and parametrized by two fixed points 
$\eta_\mu$ and $\xi_\mu$, and by a multiplier $k_\mu$, which are
related to exponentials of Schwinger proper times in field theory. 
The projective transformations $V_i'(z)$ serve to define the amplitude
off the mass shell. Finally, a word must be spent on the integration 
region, which can be deduced by studying the Schottky representation 
of Riemann surface with boundaries. Let us briefly consider the 
two--loop case for clarity. First, one uses the overall projective 
invariance of the Riemann sphere to fix, say, $\xi_1$, $\xi_2$ and 
$\eta_1$ to $\infty$, $1$ and $0$ respectively. The resulting Schottky
representation of the string world-sheet is given in Fig. 1.

\vspace{1cm}
\unitlength 0.9mm
\linethickness{0.4pt}
\begin{picture}(125.00,40.00)(10,75)
\put(80.00,100.00){\circle{8.67}}
\put(100.00,100.00){\circle{12.00}}
\put(126.00,100.00){\circle{12.00}}
\multiput(150.00,120.00)(0.11,-0.61){4}{\line(0,-1){0.61}}
\multiput(150.46,117.56)(0.10,-0.61){4}{\line(0,-1){0.61}}
\multiput(150.86,115.12)(0.11,-0.81){3}{\line(0,-1){0.81}}
\multiput(151.20,112.68)(0.09,-0.81){3}{\line(0,-1){0.81}}
\multiput(151.48,110.24)(0.11,-1.22){2}{\line(0,-1){1.22}}
\multiput(151.70,107.80)(0.08,-1.22){2}{\line(0,-1){1.22}}
\put(151.86,105.37){\line(0,-1){2.44}}
\put(151.96,102.93){\line(0,-1){2.44}}
\put(152.00,100.49){\line(0,-1){2.44}}
\put(151.98,98.05){\line(0,-1){2.44}}
\multiput(151.90,95.61)(-0.07,-1.22){2}{\line(0,-1){1.22}}
\multiput(151.77,93.17)(-0.10,-1.22){2}{\line(0,-1){1.22}}
\multiput(151.57,90.73)(-0.09,-0.81){3}{\line(0,-1){0.81}}
\multiput(151.31,88.29)(-0.11,-0.81){3}{\line(0,-1){0.81}}
\multiput(151.00,85.85)(-0.09,-0.61){4}{\line(0,-1){0.61}}
\multiput(150.62,83.41)(-0.10,-0.57){6}{\line(0,-1){0.57}}
\put(155.00,100.00){\line(-1,0){150.00}}
\multiput(10.00,80.00)(-0.11,0.61){4}{\line(0,1){0.61}}
\multiput(9.58,82.44)(-0.09,0.61){4}{\line(0,1){0.61}}
\multiput(9.21,84.88)(-0.11,0.81){3}{\line(0,1){0.81}}
\multiput(8.89,87.32)(-0.09,0.81){3}{\line(0,1){0.81}}
\multiput(8.63,89.76)(-0.11,1.22){2}{\line(0,1){1.22}}
\multiput(8.41,92.20)(-0.08,1.22){2}{\line(0,1){1.22}}
\put(8.25,94.63){\line(0,1){2.44}}
\put(8.15,97.07){\line(0,1){2.44}}
\put(8.09,99.51){\line(0,1){2.44}}
\put(8.08,101.95){\line(0,1){2.44}}
\put(8.13,104.39){\line(0,1){2.44}}
\multiput(8.23,106.83)(0.08,1.22){2}{\line(0,1){1.22}}
\multiput(8.38,109.27)(0.10,1.22){2}{\line(0,1){1.22}}
\multiput(8.59,111.71)(0.09,0.81){3}{\line(0,1){0.81}}
\multiput(8.84,114.15)(0.10,0.81){3}{\line(0,1){0.81}}
\multiput(9.15,116.59)(0.10,0.68){5}{\line(0,1){0.68}}
\put(149.17,102.00){\makebox(0,0)[cc]{$B'$}}
\put(11.50,102.00){\makebox(0,0)[cc]{$A'$}}
\put(73.60,102.00){\makebox(0,0)[cc]{$A$}}
\put(86.20,102.00){\makebox(0,0)[cc]{$B$}}
\put(92.00,102.00){\makebox(0,0)[cc]{$C$}}
\put(108.67,102.00){\makebox(0,0)[cc]{$D$}}
\put(117.50,102.00){\makebox(0,0)[cc]{$D'$}}
\put(135.20,102.00){\makebox(0,0)[cc]{$C'$}}
\put(80,108.00){\makebox(0,0)[cc]{${\cal K}_2$}}
\put(100.70,110.00){\makebox(0,0)[cc]{${\cal K}_1$}}
\put(126.70,110.00){\makebox(0,0)[cc]{${\cal K}_1'$}}
\end{picture}

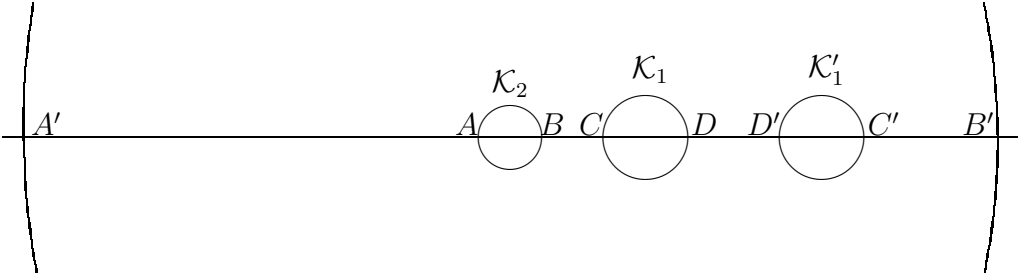
Fig. 1: The two annulus in the Schottky representation.
\vspace{0.2cm}

The two--annulus corresponds to the part of the upper--half plane
which is inside the big circle through $A'$ and $B'$, and outside 
the other circles; moreover, the points $A$, $B$, $C$ and $D$ are 
identified with $A'$, $B'$, $C'$ and $D'$ respectively. 
The segments $(AA')$ and $(DD')$ represent the two inner boundaries of
the surface, while the union of $(BC)$ and $(C'B')$ represents the 
external boundary. The positions of these points are simple functions 
of the parameters $k_1$, $k_2$ and $\eta_2$~\cite{lm_scal}. The region
of integration for the multipliers $k_i$ and for the remaining fixed 
point $\eta_2$ is determined by the request that the circles do not 
overlap, and in the field theory limit ($k_i \to 0$) it is given by 
$0 \leq \sqrt{k_1} \leq \sqrt{k_2} \leq \eta_2 \leq 1$.
The fixed point $\eta_2$ can be interpreted as the distance between 
the two loops, so that one can identify the region $\eta_2\rightarrow
1$ as related to reducible diagrams, and the region $\eta_2\rightarrow 
0$ as related to irreducible ones.

\section*{Yang--Mills vacuum diagrams}

To illustrate somewhat more practically the features of \eq{lm_hmast},
let us consider the case of the two--loop Yang--Mills vacuum
bubbles. These are obtained from \eq{lm_hmast} by setting $h = 2$,
omitting all terms involving the Green function (there are no external
legs), and expanding the measure of integration to first order in
$k_1$ and $k_2$. In this approximation, \eq{lm_hmast} becomes 
\begin{eqnarray} 
\label{lm_exp1}
A^0_2 & = & N^3 \, C_2 \, (2\pi)^d \int {dk_1\over k_1^2} {dk_2\over 
k_2^2} {d\eta_2\over (1-\eta_2)^2} \;  
\\ \nonumber &\times &
\left[1+(d-2)(k_1+k_2)+\left((d-2)^2+d\,(1-\eta_2)^2{1+\eta_2^
2 \over \eta_2^2}\right)k_1 k_2\right]
\\ \nonumber &\times &
\Bigg[\ln k_1\ln k_2 - \ln^2\eta_2 + {2\,(1-\eta_2)^2\over 
\eta_2} (k_1\ln k_1 + k_2\ln k_2) +
\\ \nonumber & &
+ {4\,(1-\eta_2)^4\over \eta_2^2}\left(1 + {1+\eta_2 \over 1- \eta_2}
\ln\eta_2 \right) k_1 k_2\Bigg]^{-d/2}~~~. 
\end{eqnarray}

There are three singular integration regions that contribute to the
amplitude for massless states. The simplest is the  region $\eta_2 \to
0$, in which the massless states are selected by extracting from the 
integrand in \eq{lm_exp1} the terms proportional to $k_1^{-1} k_2^{-1}
\eta_2^{-1}$. Here it is convenient 
to introduce the variables $q_1 = k_2/\eta_2$, $q_2 = k_1/\eta_2$, 
$q_3 = \eta_2$, which are directly related to the field theory proper times 
by $t_i=-\a'\ln q_{i}$. The region of integration now takes the form 
$0\leq q_1\leq q_2\leq q_3\leq 1$, and the term that survives in the 
limit $\a'\rightarrow 0$ is given by
\begin{equation} 
\label{lm_asym}
A^0_2\big|_{q_3\rightarrow 0} = \frac{g_d^2}{(4\pi)^d} N^3 d (d-2) 
\int\limits_0^\infty 
dt_2 \int\limits_0^{t_2} dt_1 \int\limits_0^{t_1} dt_3
{t_1+t_2+2t_3 \over (t_1 t_2+t_1 t_3+t_2 t_3)^{1+d/2}}~.
\end{equation}
Notice that the integrand is not symmetric in the exchange of the
three proper times, so that the limits of integration cannot be
extended to $\infty$ at this stage. This desease is treated by
including the second singular region, which can be parametrized as
$q_3 = y q_2$. In this region $q_3$ and $q_2$ go to zero at the same
speed (they are not strongly ordered), so that we do not associate 
any proper time to the variable $y$, which is kept finite.
This region contributes
\beq
\label{lm_asym2}
A^0_2\big|_{q_3\rightarrow q_2} = -
\frac{g_d^2}{(4\pi)^d} N^3 2 (d-2) \int_0^\infty 
dt_2 \int_0^{t_2} dt_1 (2 t_1 t_2+t_2^2)^{-d/2}~,
\eeq
where the integrand can be rewritten as
\beq
\label{lm_id}
(2 t_1 t_2+t_2^2)^{-d/2} = (t_1 t_2)^{-d/2} + \int_0^{t_1} dt_3 
\frac{t_1+t_2}{(t_1 t_2 + t_1 t_3 + t_2 t_3)^{1 + d/2}}~.
\eeq
\eq{lm_asym2} thus symmetrizes \eq{lm_asym}, as well as contributing 
to a contact interaction. 

Finally, in the region $\eta_2\rightarrow 1$, we should extract from
the integrand of \eq{lm_exp1} the terms proportional to $k_1^{-1}
k_2^{-1} (1-\eta_2)^{-1}$, and introduce the proper times $t_i=-\a'\ln
k_{i}$, $t_3=-\a' \ln(1-\eta_2)$. It can be checked that no such terms
survive in the limit $\a'\rightarrow 0$, which is as expected, since 
the reducible diagram we are considering is zero in Yang--Mills
theory. There is however a contact interaction term, leftover from 
tachyon exchange in the limit $\eta_2\rightarrow 1$ (such terms, 
contributing to diagrams with four--gluon vertices, were present 
already at one loop~\cite{lm_us}). This is obtained by isolating the
term in \eq{lm_exp1} that is proportional to $k_1^{-1} k_2^{-1}
(1-\eta_2)^{-2}$, and requiring that the integrand be independent of
$\eta_2$, except for the tachyon double pole. The double pole is then 
regularized using a $\zeta$-function regularization, amounting to
the substitution $\int_0 d x/x^2 \to  - 1/2$. The remaining
integral over the proper times $t_1$ and $t_2$ is then
\beq
\label{lm_tach}
A^0_2\big|_{\eta_2\rightarrow 1} = - \frac{g_d^2}{(4\pi)^d}
N^3 \frac{(d-2)^2}{4} \int\limits_0^\infty dt_1\, dt_2 (t_1 t_2)^{-d/2}~.
\eeq

It is easy to check that the sum of Eqs.~(\ref{lm_tach}), (\ref{lm_asym}) 
and (\ref{lm_asym2}) correctly reproduces the sum of the corresponding 
Feynman diagrams in Yang-Mills theory. Notice that the organization of
the different terms arising from string theory is very different from
what might have been expected in field theory.

\section*{A two--loop scalar diagram}

We will now consider a slightly more complicated diagram, a
contribution to the two--point function with both external legs on the
same propagator. Since we consider scalar particles (tachyons of the
bosonic string), we need the expansions of the various functions in
\eq{lm_hmast} only to leading order. The measure can be read off from
\eq{lm_exp1}, while the Green function is given by
\beqa
{\cal{G}}^{(2)} (z_1 , z_2) & = & \log ( z_1 - z_2 ) + \frac{1}{2}
\left[ \log k_1  \log k_2 -  \log^2 S \right]^{-1} \label{lm_green} \\
& \times & \left[ \log^2 T \log k_2 + \log^2 U \log k_1  - 2 \log T \log U
\log S \right]~~,   \nonumber
\eeqa
where 
\beq
S = \eta_2~, ~~~
T = \frac{z_2 }{z_1}~,~~~
U  = \frac{(z_2 - \eta_2 ) (z_1 - 1)}{(z_1 - \eta_2 )
(z_2 -1 )}~~.
\label{lm_stu}
\eeq
The projective transformations $V_i(z)$, to leading order in the
multipliers, are given in~\cite{lm_scal}.

Contributions to scalar irreducible diagrams can only come from the
region $\eta_2\rightarrow 0$; moreover, since we are interested in
planar amplitudes, both states must lie on the same boundary of the
two--annulus. These observations completely determine the region of 
integration of the variables in \eq{lm_hmast}, of which we must now
consider only the first line. Furthermore, it is possible to identify
the integration region associated with each field theory diagram
already at the string level. In the field theory limit, the
world-sheet becomes a graph, and each world--sheet boundary
degenerates into the union of two distinct propagators, on each of
which the external legs may be attached. One can then
determine which integration regions for the punctures $z_i$ are
related to insertions on the various propagators.
For example, if a puncture $z$ lies on the internal boundary
represented by the segment $(AA')$, in the corresponding field theory 
diagram it will always be attached to the first loop, but will be 
emitted from the internal propagator if $z\in[-1,-\eta_2]$, or from 
the external one if $z\in[-\eta_2,A]$ and $z\in[A',-1]$.

In order to show that this identification is correct, let us calculate 
the two loops diagram with both external states in the region 
$[-1,-\eta_2]$. Performing the field theory limit, we will 
transform the tachyon into a state of real mass rewriting the quadratic 
poles $x^{-2} \rightarrow x^{-1-a}=x^{-1}\exp [m^2 \a' \log x]$, as
done in Ref.~\cite{lm_scal}.

In the relevant region ($-1<<z_1<<z_2<<-\eta_2$) the first line of 
\eq{lm_hmast} becomes
\bea \nonumber
A^2_2 & = & \!\! \frac{N^2}{2^{10}} {g^4\over 
(4\pi)^d} \int_0^\infty\!\!\! dt_1\int_0^{t_1}\!\!\! dt_2
\int_0^{t_2}\!\!\! dt_3{\rm e}^{-m^2(t_1+t_2+t_3)}
(t_1 t_2+t_1 t_3+t_2 t_3)^{-d/2}\\ &\times&\label{lm_fish}
\int_0^{t_3}\!\! dt_4\int_0^{t_4}\!\! dt_5 ~\exp\left\{p_1\cdot 
p_2\left[(t_4-t_5)-{(t_1+t_2)(t_4-t_5)^2\over (t_1 t_2+t_1 t_3+t_2 t_3)}
\right]\right\}
\eeqa
where new Schwinger parameters ($t_4=-\alpha\ln |z_1|$ and
$t_5=-\alpha\ln |z_2|$) have been introduced. 
The region of integration in \eq{lm_fish} is naturally understandable 
in a first quantized framework: the proper time related to the
punctures cannot exceed the lenght of the propagator on which they are
inserted. Identical contributions come from the regions where the
order of $z_1$ and $z_2$ is changed, and where the two external states
are attached to the second boundary, represented by the segment
$(DD')$. When the particles are emitted by the two other propagators, 
one gets contributions that are simply a cyclic permutation of $t_1$, 
$t_2$ and $t_3$ in \eq{lm_fish}. Taking into account all relevant
regions, one gets a completely symmetric expression and can perform
the integration over $t_1$, $t_2$ and $t_3$ independently, introducing
a factor of $1/3!$. With the natural change of variables $t_{1,2} = 
x_{1,2}$, $t_3 = x_3+x_4+x_5$, $t_4 = x_4+x_5$ and $t_5 = x_5$, one 
obtains the final expression 
\beqa 
A^2_2 & = & \frac{N^2}{2^{9}}\,{g^4\over (4\pi)^d}
\int_0^\infty\!\! \prod_{i=1}^5dx_i~~ \Delta^{-d/2} 
\label{lm_fish1} ~{\rm e}^{-m^2\sum_{i=1}^5 x_i} \\ &\times & \nonumber
\exp{\left\{p_1\cdot p_2\left[\Delta^{-1}\Big(x_4 (x_1 (x_2+x_3+x_5)
+ x_2(x_3+x_5)
\Big)\right]\right\}}~~,
\eeqa
where $\Delta = x_1 x_2 + (x_3+x_4+x_5) (x_1 + x_2)$. This is the
correctly normalized expression for the correspnding diagram in field
theory. 

Similar calculations with gluons are in progress. For the two-point
function, which must be calculated off-shell, the main remaining
difficulty is the extension of the expression given in~\cite{lm_scal}
for the projective transformations $V_i(z)$ beyond leading order in the
multipliers. This difficulty is however irrelevant for higher--point
amplitudes, which can be calculated on shell and are independent of
the $V_i$'s.

\end{document}